# US Energy-Related Greenhouse Gas Emissions in the Absence of Federal Climate Policy

Hadi Eshraghi[1], Anderson Rodrigo de Queiroz[2], Joseph F. DeCarolis[1,*]

[1] Department of Civil, Construction, & Environmental Engineering, NC State University, Raleigh, NC, 27695
[2] Decision Sciences Department, School of Business, NC Central University, Durham, NC, 27707.
[*]Corresponding author email: jdecarolis@ncsu.edu

**ABSTRACT**

The planned US withdrawal from the Paris Agreement as well as uncertainty about federal climate policy have raised questions about the country's future emissions trajectory. Our model-based analysis accounts for uncertainty in fuel prices and energy technology capital costs and suggests that market forces are likely to keep US energy-related greenhouse gas emissions relatively flat or produce modest reductions: in the absence of new federal policy, 2040 greenhouse gas emissions range from +10% to -23% of the baseline estimate. Natural gas versus coal utilization in the electric sector represents a key tradeoff, particularly under conservative assumptions about future technology innovation. The lowest emissions scenarios are produced when the cost of natural gas and electric vehicles decline while coal and oil prices increase relative to the baseline.

**INTRODUCTION**

The US is the second largest energy-related greenhouse gas emitter [1], and therefore critical to global efforts to mitigate climate change. The US intends to formally withdraw from the Paris Agreement, and key policies aimed at curbing greenhouse gas emissions – in particular the Clean Power Plan and revised CAFE standards – face a highly uncertain fate. Inaction on the federal level is tempered by state-level action, including California's SB32 [2], the Regional Greenhouse Gas Initiative (RGGI) covering 9 northeastern states [3], and renewable portfolio standards active in over 35 states [4]. In addition to federal and state policy, market forces will play a critical role in shaping the future US energy system over the next several decades. Reasons for optimism include low natural gas prices [5] as well as dramatic drops in the cost of solar photovoltaics [6] and lithium ion batteries used for grid storage and electric vehicles [7]. While prevailing market forces are likely to temper unconstrained greenhouse gas emissions at the national level, the US will eventually need to produce deep emissions reductions in order to avoid the worst effects of climate change. The US had previously acknowledged the need for such reductions. For example, the US nationally determined contribution (NDC) under the Paris Agreement is 26-28% below 2005 levels by 2025, and the Mid-Century Strategy suggests an 80% reduction below 2005 levels by 2050 [8].

---

[1] Corresponding author: jfdecaro@ncsu.edu

Given the anticipated lack of near-term federal action to address climate change, it is critical to evaluate potential baseline emissions scenarios in the absence of federal climate policy. In addition, careful model-based analysis of baseline scenarios can help inform discussions regarding the timing and structure of future climate and energy policy. The US Energy Information Administration (EIA) utilizes the National Energy Modeling System (NEMS) to produce the Annual Energy Outlook (AEO) [9]. The AEO includes a Base Case as well as several side cases that typically focus on variations in economic growth, fuel resource cost and availability, and rates of technology innovation. For example, AEO 2017 includes a total of seven cases that are repeated with and without implementation of the EPA Clean Power Plan [9]. While these internally consistent scenarios provide a sketch of potential midterm energy futures, they belie the underlying market uncertainty that could push the US energy system in different directions in the absence of new policy. Several other recent modeling efforts have projected US energy technology deployment and greenhouse gas emissions, but generally focus on scenarios under proposed or hypothetical federal policy and use a limited number of scenarios to address parametric uncertainty [10,11,12,13,14,15,16,17,18].

In this analysis, we utilize Tools for Energy Model Optimization and Analysis (Temoa) [19], an open source, publicly available energy system optimization model (ESOM) to examine a large set of baseline US energy futures through 2040. Our objective is to rigorously explore the future decision landscape and quantify greenhouse gas (GHG) emissions in a future where energy system changes are driven by market forces rather than top-down federal policy. We employ a sensitivity technique called the Method of Morris [20,21] to rank order the input parameters that produce the largest effect on emissions. We then incorporate the ten most sensitive parameters into a suite of Monte Carlo simulations that indicate how US energy-related GHG emissions may change under different future assumptions. The full set of results are used to identify plausible combinations of assumptions that can lead to either very high or low emissions, which can inform our understanding of future baseline emissions and suggest pathways to lower emissions in the absence of new federal policy.

## MODEL AND DATA

The analysis is performed with an open source energy system optimization model (ESOM) called Temoa, which operates on a single region input database representing the continental United States. The model source code and data are publicly available through GitHub [22], and an exact copy of the files used to produce this analysis are archived through zenodo [23]. Key features of the model and input dataset are described here, with additional information provided in Section 1 of the SI.

**Tools for Energy Model Optimization and Analysis (Temoa).** Temoa [19] is an open source, bottom-up ESOM, similar to MARKAL/TIMES [24], OSeMOSYS [25] and MESSAGE [26]. Temoa employs linear optimization to generate the least-cost pathway for energy system development. The model objective function minimizes the system-wide present cost of energy provision over a user-specified time horizon by optimizing the installation and utilization of energy technologies across the system. Technologies in Temoa are explicitly defined by a set of engineering-economic parameters (e.g., capital costs, operations and maintenance costs, conversion efficiencies) and are linked together in an energy system network through a flow of energy commodities. Model constraints enforce rules governing energy system performance, and user-defined constraints can be added to represent limits on technology expansion, fuel availability, and system-wide emissions. The model formulation is detailed in Hunter et al. [19] and the Temoa source code is publicly available on Github [22]. Since the model formulation evolves over time, revised model documentation can be found on the project website [27].

**Input data:** The input database used in this analysis is largely drawn from the EPA MARKAL database [28] and represents the US as a single region. The time horizon extends from 2015 to 2040, with 5-year time periods. For example, the 2015 period covers the years 2015 to 2019. The results for each year within a given time period are assumed to be identical. Temporal variation in renewable resource supply and end-use demands is captured through representation of three seasons (summer, winter, intermediate) and four times of day (am, pm, peak, night). Fuel price trajectories are drawn from the Annual Energy Outlook (AEO) [9] and specified exogenously. While assuming a fixed fuel price trajectory does not capture demand-price feedbacks, it simplifies the execution and interpretation of the sensitivity analysis. The model tracks emissions of $CO_2$, $NO_x$ and $SO_2$ as well as $CH_4$ leakage rates from natural gas systems. We assume a methane leakage rate equivalent to 1.4% of total natural gas delivered [29], which is lower than both NETL [30] and EDF [31] estimates of 1.6 and 1.65%, respectively. Given the ability to mitigate methane leakage and the multi-decadal timescale of our analysis, use of the EPA estimate was deemed appropriate. Methane emissions are transformed into $CO_2$-equivalents using a global warming potential (GWP) of 25 [29]. This GWP value complies with the international inventory reporting guideline under the United Nations Framework Convention on Climate Change [29]. The input database and baseline assumptions are detailed in Supporting Information, and a brief sectoral description of the input dataset is provided in Table 1.

The Temoa baseline scenario is designed to be conservative. The baseline assumes that the Clean Power Plan is not implemented, and does not include California's cap-and-trade system or RGGI. The baseline includes the aggregate effect of state-level renewable portfolio standards as well as prevailing tax incentives, including the production tax credit for wind[32] and a 10% tax credit on utility scale solar PV throughout the time horizon[33]. To orient our baseline to a familiar projection, our input assumptions draw heavily on the AEO[9] and Assumptions to the AEO[34]. The Temoa baseline results are compared with AEO in Supporting Information Figures S2-S7.

**Table 1.** Sectoral-level detail in the Temoa input database

| Sector | Description |
|---|---|
| Fuel Supply | Fuel prices are specified exogenously. Baseline projections are drawn from the 2017 Annual Energy Outlook [9]. There is no limit on fuel availability except for biofuel use in the transportation sector [35]. |
| Electric | The electric sector includes 34 generating technologies. Air pollution control retrofits for coal include low NOx burners, selective catalytic reduction, selective non-catalytic reduction, and flue gas desulfurization. Costs and performance characteristics are largely drawn from the EPA U.S. nine-region MARKAL database [28], and existing capacity estimates are drawn from the US EIA [9]. |
| Transportation | The transportation sector is divided into four modes: road, rail, air, and water. Road transport is modeled with greater detail by dividing it into three subsectors: light duty transportation, heavy duty transportation, and off-highway transportation. The light duty sector includes 6 size classes and 9 different vehicle technologies. Data is largely drawn from the EPA U.S. nine-region MARKAL database [28]. |
| Industrial | Given the high degree of heterogeneity in the industrial sector, it is modeled simplistically as a set of fuel share constraints that are calibrated to the 2017 Annual Energy Outlook [9]. |
| Commercial | The commercial sector includes the following end-use demands: space heating, space cooling, water heating, refrigeration, lighting, cooking, and ventilation. A total of 83 demand technologies are included to meet these end-use demands. Data is largely drawn from the EPA U.S. nine-region MARKAL database [28]. |
| Residential | The residential sector includes the following end-use demands: space heating, space cooling, water heating, freezing, refrigeration, lighting, cooking, and appliances. A total of 69 demand technologies are included to meet these end-use demands. Data is largely drawn from the EPA U.S. nine-region MARKAL database [28]. |

## ANALYSIS FRAMEWORK

Our methodological approach shares common elements with previous work. For example, we utilize large scale scenario generation and cluster analysis similar to the Robust Decision Making (RDM) approach [36,37]; however, we are not attempting to identify a policy strategy. In addition, recent studies have used ESOMs to generate a large ensemble of near optimal scenarios to derive policy relevant insights, but such work has focused on European applications [38,39,40]. Below we describe each element of our framework in turn.

**Method of Morris.** Following work by Usher [41], we utilize a global sensitivity method called Method of Morris [20,21] to identify the model inputs that produce the largest effect on cumulative GHG emissions over the model time horizon. The method produces a reliable sensitivity measure with a minimum number of runs and can handle a large number of uncertain parameters, making it suitable for use with data-intensive ESOMs [41]. We consider price variation in 6 different fuels and 35 technology-specific capital costs. (See Supporting Information for additional details on the Method of Morris formulation and problem setup used in this analysis.) For simplicity, each parameter is varied within a range representing ±20% of its baseline value rather than trying to identify specific ranges for each parameter separately, which are subject to considerable future uncertainty.

**Monte Carlo Simulation.** Next, we perform a Monte Carlo simulation where we consider variation in the ten most sensitive parameters selected from the Method of Morris analysis. Our objective is to quantify how variation in the ten most sensitive techno-economic parameters can affect the resultant range in GHG emissions. Given the high uncertainty associated with these future parameter values, we do not attempt to quantify different ranges, probability distributions, or correlations between parameters. Rather, a uniform distribution and range is assumed for each parameter, similar to other studies [40,42,43,44]. As a result, the full set of model results indicate the range of future emissions pathways and suggest possible outcomes, but should not be interpreted probabilistically. When investigating low emissions outcomes relying on specific combinations of realized parameter values, we consider the plausibility of those parameter combinations ex post. The required number of model runs for the Monte Carlo simulation is assumed independent of the number of uncertain inputs [45]; 1000 runs are conducted within the simulation. To minimize the computational time, we create an embarrassingly parallel [46] implementation of the framework. The model runs are parallelized using the "joblib" python library [47]. We run the model using a workstation containing two Multi-Core Intel Xeon E5-2600 series processors, representing a total of 12 compute cores.

**K-means clustering.** Rather than examine the raw set of 1000 model runs, we employ k-means clustering to examine a limited number of representative points. The k-means algorithm partitions the dataset by creating groups or clusters with similar features. The algorithm minimizes the Euclidean

 

distance between the centroids of each cluster, where each cluster consists of centroid values representing the 10 uncertain input parameters plus cumulative emissions (see Supporting Information for more details). We separate the data into ten clusters, which provide enough points to identify relationships between input values and the resultant level of cumulative $CO_2$ emissions. Larger numbers of clusters were tested, but the configuration of centroids did not yield additional insights.

The $k$-means clustering algorithm is a well-established methodology applied to separate datasets into homogenous groups of observations. The method was first developed by Lloyd [48] and has been widely used as a non-hierarchical clustering approach. Other methods such as principal component analysis [49], hierarchical and other non-hierarchical clustering methods [50,51], and supervised and unsupervised learning algorithms [51] could also be used for our purpose. However, in this work we make use of the $k$-means method for clustering due to its simplicity, efficiency, and successful application in several areas of the literature [52].

**Uncertainty Cases.** We develop three different cases to represent different levels of future uncertainty, and repeat the Monte Carlo simulation, consisting of 1000 model runs, for each case. We refer to the first case as 'Stable World,' which denotes a relatively stable future in which the ten most sensitive parameters selected from Method of Morris vary within ±20% of their baseline values. The second case, 'Uncertain Fuels,' allows natural gas and oil prices to vary within ±80% of their baseline values, consistent with their longer-term historical range over the last 50 years [53,54]. The remaining eight parameters in the Uncertain Fuels case vary within ±20% of their baseline values, as in the Stable World case. The third case, 'Uncertain World,' allows natural gas and oil prices to vary within ±80% of their base value, while the other eight uncertain input factors vary within ±40% of their baseline values.

## RESULTS AND DISCUSSION

The presentation of results follows the order described in the Analysis Framework section. We begin by presenting results from the Method of Morris sensitivity analysis, followed by the Monte Carlo simulations associated with each of the three uncertainty cases. The raw Monte Carlo results are used to examine the range of future cumulative emissions and the role that natural gas prices play in determining emissions. Finally, we present results from $k$-means clustering to assess how variations in technology cost and fuel prices lead to different emissions outcomes.

**Identifying key sensitivities.** The Method of Morris results (Figure 1) indicate that natural gas prices have the largest overall effect on cumulative GHG emissions. In the electric sector, coal prices and capital costs for solar photovoltaics, wind, and combined-cycle gas turbines also have a measurable effect on total emissions. The inclusion of capital costs associated with battery electric vehicles, conventional gasoline vehicles, and diesel vehicles indicate that the light duty vehicle sector can also have an effect on emissions. Below the tenth most sensitive parameter (heat pump capital cost), the average effect on cumulative GHG emissions is less than 0.25% of the base case cumulative emissions. In general, the small relative changes in cumulative emissions reflect inertia in the energy system: a change in any single parameter takes time to reach its full effect on technology deployment and has a limited effect across the system.

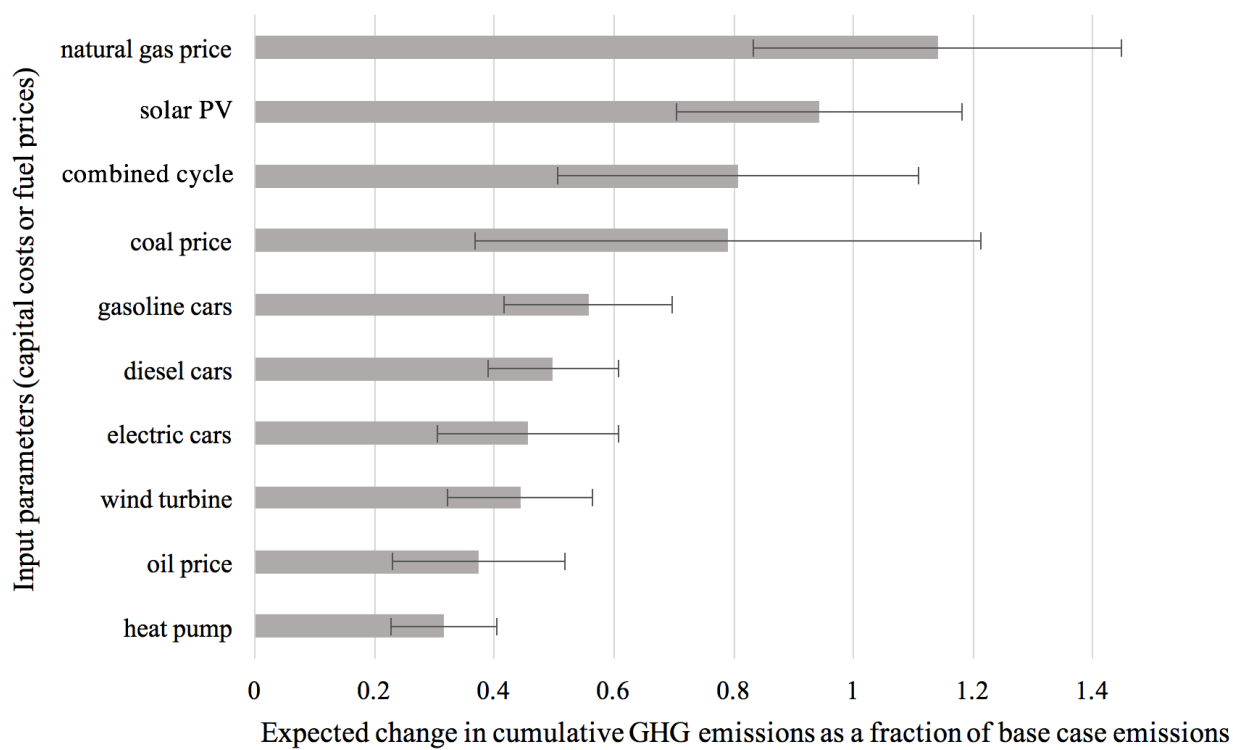

**Figure 1.** Method of Morris results indicating the ten input parameters that produce the largest effect on cumulative GHG emissions (2015-2040), ranked from largest to smallest effect. Parameters labeled "price" represent fuel prices, while all others represent capital costs. The horizontal axis indicates the magnitude of the expected change in cumulative GHG emissions relative to the baseline value. Each input parameter is tested at 25 distinct values over a range representing ±20% of its baseline value. The length of the bar indicates the average effect, while the error bars indicate the 95% confidence intervals.

We repeated the Method of Morris analysis with a ±40% input parameter range and found that it generates the same top ten parameters as shown in Figure 1; however, oil price rises to the second rank while the relative order of the other inputs stays the same.

**Baseline GHG emissions under future uncertainty.** The ten parameters with highest sensitivity (Figure 1) are selected for inclusion in a suite of Monte Carlo simulations that indicate how US energy-related GHG emissions may change under different future assumptions. The distribution of cumulative GHG emissions from the three cases is shown in Figure 2 where kernel density estimation [50] is employed to smooth out the raw histogram results.

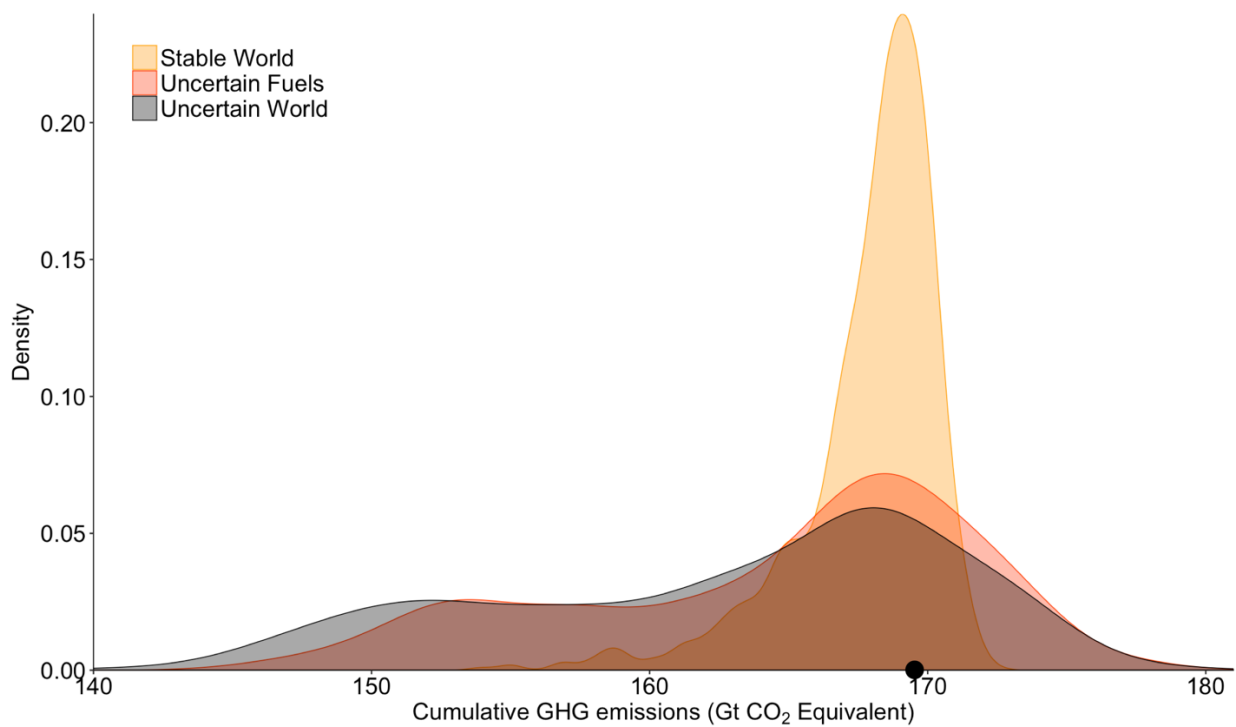

**Figure 2.** Kernel density estimates of cumulative GHG emissions from 2015-2040 for three cases: Stable World, Uncertain Fuels, Uncertain World. The modeled baseline GHG emissions are estimated to be 169 Gtonnes of $CO_2$ equivalent, represented by the black dot on the horizontal axis. Larger ranges in input parameters produce large ranges in cumulative GHG emissions, with results skewed towards cumulative emissions below the baseline value.

In the Stable World case, the distribution of cumulative GHG emissions are clustered around the baseline scenario (169 $GtCO_2e$), with a range extending to a minimum emissions level of 153 $GtCO_2e$. By comparison, both the Uncertain Fuels and Uncertain World cases exhibit a wider range in cumulative GHG emissions than Stable Word, but both are skewed towards lower emissions. Thus, allowing a wider range in fuel prices (±80%) flattens the distribution of cumulative emissions and increases the proportion of scenarios with emissions lower than the baseline. Moving from Uncertain Fuels to Uncertain World increases the highest emissions scenario by 1% and decreases the lowest emissions scenario by 3.2% relative to the cumulative emissions level in the baseline scenario. Overall, Figure 2 indicates that wider input ranges related to fuel costs and technology investment costs increase the proportion of emissions scenarios below the baseline. For reference, our baseline cumulative GHG emissions are 6.2% higher than the AEO reference case without the Clean Power Plan [9]. Part of this discrepancy is due to our consideration of $CO_2$-equivalent emissions from methane leakage during natural gas production, processing and transport, which AEO does not report. If only $CO_2$ emissions are compared, the difference is 3.2%. Across all modeled scenarios, methane leakage ranges from 1.6% to 4.1% of total $CO_2e$ emissions.

The GHG emissions trajectories associated with the three cases are presented in Figure 3 and compared with the Mid-Century Strategy (MCS) for deep decarbonization [8]. The MCS outlines a path for the US to meet its commitments under the Paris Accord and ultimately achieve an 80% $CO_2$ reduction below the 2005 level by 2050.

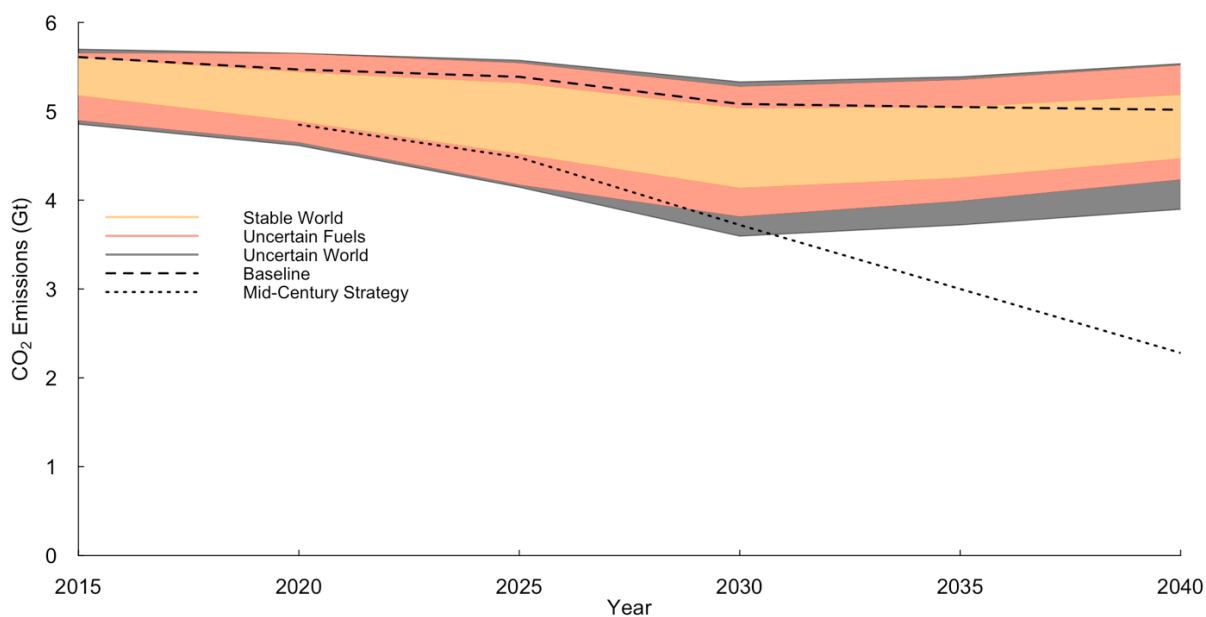

**Figure 3.** Ranges of projected $CO_2$ emission pathways in the three modeled cases, with the baseline emissions scenario and the Mid-Century Strategy (MCS) included for reference. All three cases could result in emission pathways significantly lower than the base case. In the first model time period (2015-2019), technology capacity remains fixed except for new wind and solar, which have been experiencing rapid year-over-year growth. Thus, emissions variations in the first time period are due to differences in the utilization of existing capacity as well as new installed renewable capacity. See Supporting Information for further discussion of the baseline scenario results.

Figure 3 indicates that it may be possible to meet the US 2025 commitments in the absence of federal policy; however, market forces alone are not enough to sustain the emissions reductions prescribed by the MCS post-2025.

**The effect of fuel prices in the power sector on GHG emissions.** In addition to total GHG emissions, we examine the underlying trends in technology deployment that drive the emissions shown in Figures 2 and 3. Since the Method of Morris results indicated that emissions are highly sensitive to natural gas and coal prices, we plot cumulative GHG emissions versus the average ratio of natural gas to coal prices across all model time periods (Figure 4). In the Stable World case (Figure 4a), there is a linear increase in emissions as the natural gas price increases relative to coal, which is due to the direct substitution of natural gas with coal to produce baseload electricity. Around a price ratio of approximately 1.8, however, the cumulative GHG emissions reach a plateau because baseload electricity production from coal reaches a maximum. At low price ratios, the variation in emissions at a given fixed price ratio is largely explained by variation in the capital cost of advanced natural gas combined cycle capacity. However, at higher price ratios above 1.8, the variability in cumulative emissions increases as variations in other input parameter values begin to exert their influence under high natural gas prices.

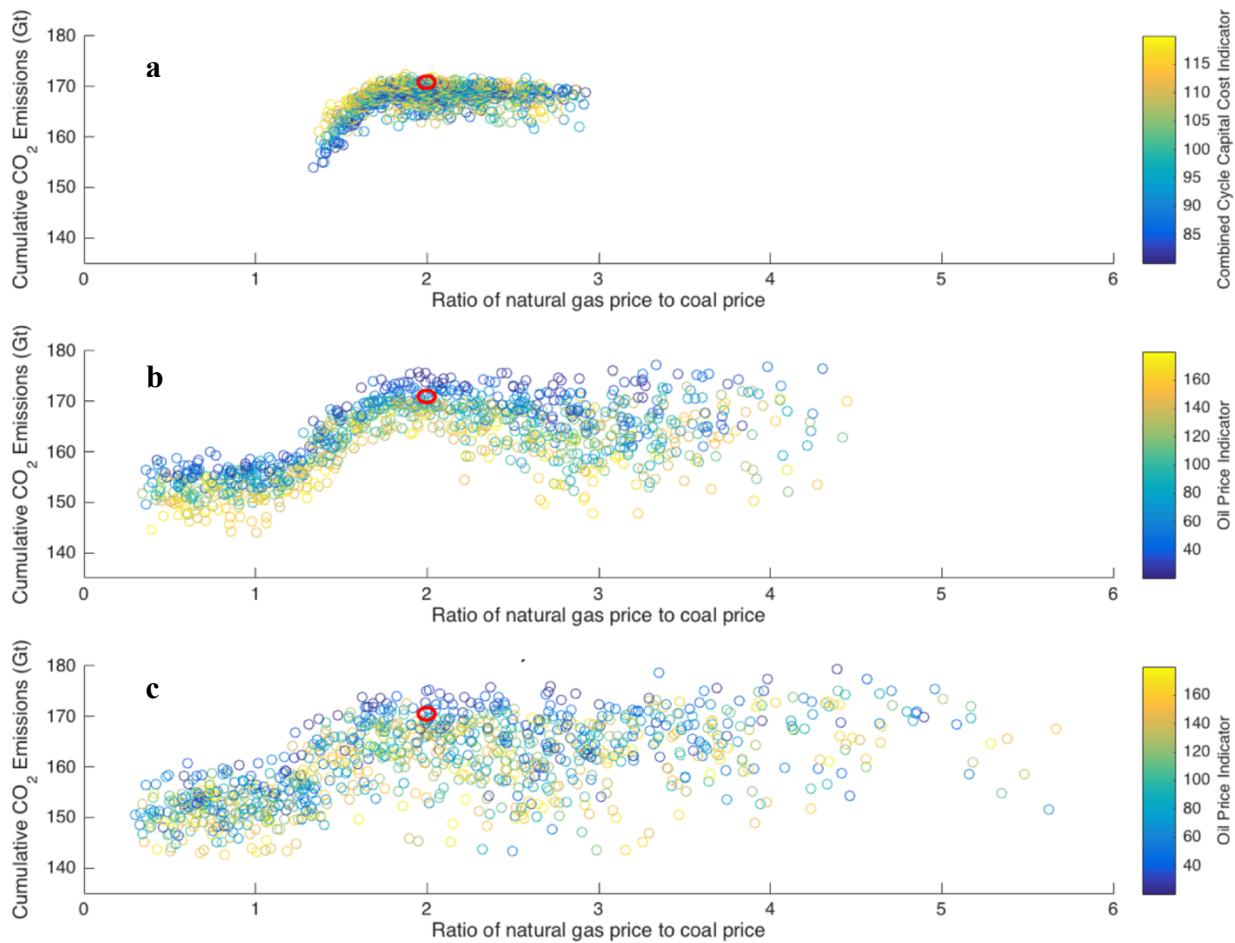

**Figure 4.** Cumulative GHG emissions versus the ratio of natural gas to coal prices. Each subplot represents the full set of 1000 runs associated with each case: (a) Stable World, (b) Uncertain Fuels, and (c) Uncertain World. In each case, the red circle represents the baseline projection. Each point in the Stable World case is colored by the capital cost of combined-cycle natural gas turbines, while points in the other two cases are colored by the oil price. These factors help explain the variability in cumulative GHG emissions at a given fuel price ratio. The color bar indicates the scalar value used to adjust the input parameter value in the Monte Carlo simulation.

In the Uncertain Fuels case (Figure 4b), coal and natural gas prices still largely explain cumulative emissions when the price ratio is below 1.8, as in the Stable World case. However, the wider range associated with input natural gas and oil prices in the Uncertain Fuels case leads to a wider range in cumulative GHG emissions. The maximum variation in GHG emissions at a given fuel price ratio is approximately 33 GtCO₂e in the Uncertain Fuels case, and 18.4 GtCO₂e in the Stable World case. While the spread in cumulative emissions increases in the Uncertain Fuels case, it is largely skewed towards lower emissions. At a given natural gas to coal price ratio, oil prices help explain the variation in cumulative emissions, particularly at price ratios less than two.

In the Uncertain World case (Figure 4c), the variability in cumulative emissions as a function of fuel price ratio further increases because other input parameters play a larger role in determining emissions. Compared with the Uncertain Fuels case, oil prices are not as clearly correlated with cumulative emissions at a given price ratio. Emissions in all three cases are skewed towards lower values. In addition, there is a fairly consistent emissions ceiling; cumulative emissions do not exceed 180 $GtCO_2e$ in any of the three cases.

**The effect of all uncertain inputs on GHG emissions.** Figure 4 indicates that the cumulative GHG emissions are strongly influenced by input parameters other than natural gas and coal prices in the Uncertain Fuels and Uncertain World cases. *K*-means clustering is applied to Monte Carlo results to condense the full set of 1000 runs from each case into a more manageable 10 clusters, which can be used to identify other key input parameters influencing cumulative emissions. Each of the ten clusters is defined by ten centroids representing the input parameter scaling factors used in the Monte Carlo simulation and another centroid representing cumulative GHG emissions. The centroids are extracted from their clusters, grouped by input parameter, and plotted versus the associated cumulative emissions in Figure 5. Parameters that demonstrate a monotonic relationship with cumulative emissions and a wider spread in centroid values suggest a stronger effect on the emissions outcome.

Spearman rank correlation coefficients are used to quantify the relationship between the centroid values and associated cumulative emissions. Spearman coefficients quantify the correlation between parameter value ranks, and are thus an appropriate choice because they measure the degree of monotonicity between variables and do not require a linear relationship. High Spearman coefficients with low p-values ($<0.05$) indicate that changing a given input parameter produces a consistent directional change in emissions. The capital costs of solar PV, wind, electric vehicles, and heat pumps as well as natural gas, coal, and oil prices have high Spearman coefficients (>0.6) and low p-values (<0.05) in at least one of the Uncertain Fuels and Uncertain World cases. Coal and oil prices exhibit negative correlation, while renewable and heat pump capital costs as well as natural gas prices show positive correlation with emissions.

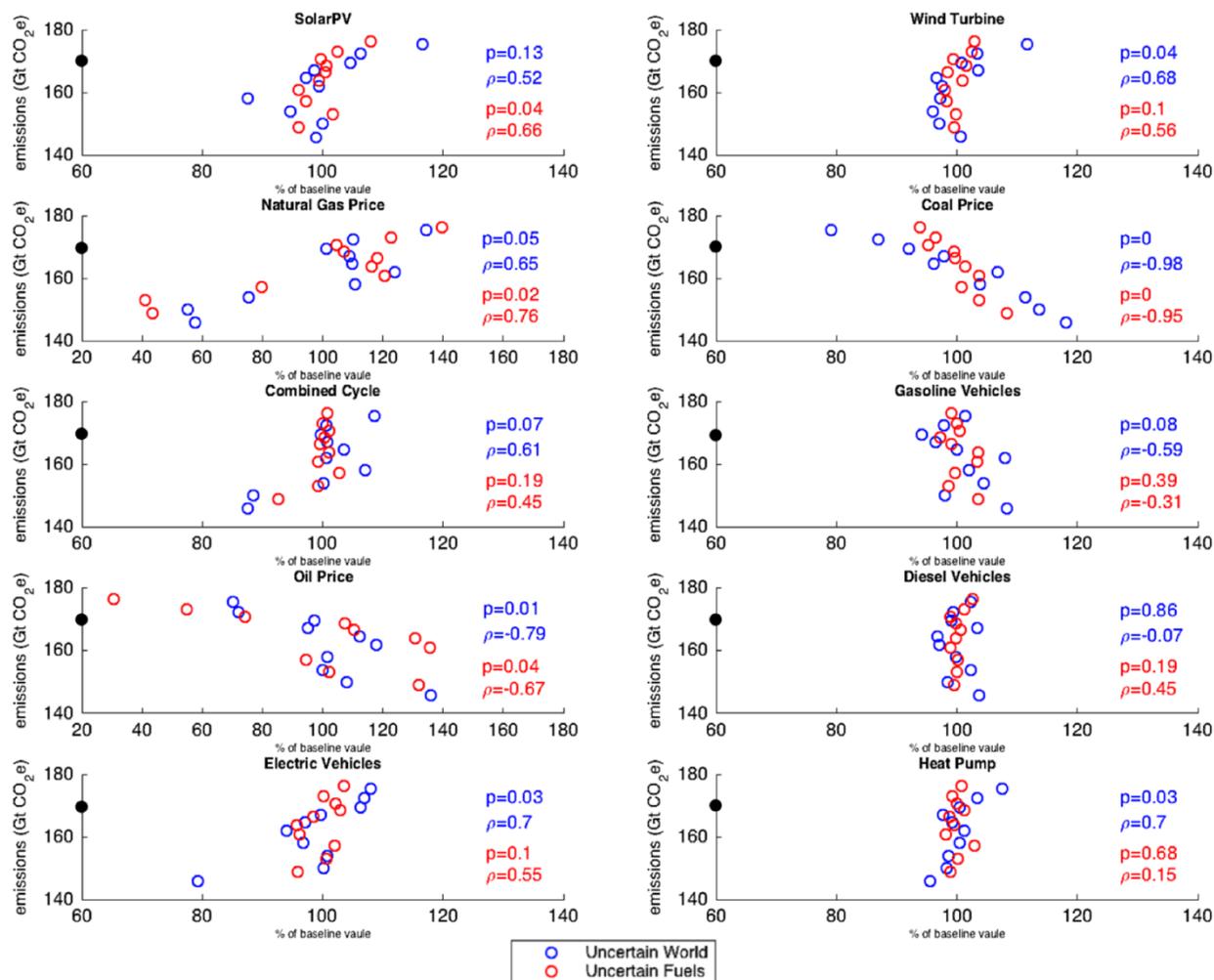

**Figure 5.** Centroid values associated with uncertain inputs (x-axis) versus cumulative GHG emissions (y-axis). The centroid values represent scaling values, which are expressed as a fraction of the assumed baseline value. X-axis ranges correspond to the allowable ranges in the Uncertain Fuels and Uncertain World cases. The Spearman rank correlation coefficients ($\rho$) and p-values help to identify the degree of monotonicity between each input and emissions. Cumulative baseline emissions are shown by the black dot on the y-axis.

Although heat pump capital cost exhibited a low coefficient of variation (3%) in the Uncertain World case, we investigated the raw scenario results further and found that it had little effect on cumulative emissions.

**Assessment of the highest and lowest emissions outcomes.** The cluster results can also be used to identify the parameter combinations that produce the highest and lowest emissions outcomes, which can inform future policy discussions. Clustering analysis is applied separately to the 50 model runs in both the Uncertain Fuels and Uncertain World cases that produce the highest and lowest 5% cumulative GHG emissions (Figure 6). In Figure 6, centroids are grouped by cluster to demonstrate how a particular set of centroids comprising a cluster produce a given emissions outcome. We consider the six input parameters with high Spearman correlation coefficients (>0.6) that are statistically significant at the 5% level in either the Uncertain Fuels and Uncertain World cases and whose centroids have a coefficient of variation greater than 10%. Two clusters per case and emissions level (high or low) are generated; more clusters tended to produce redundant results.

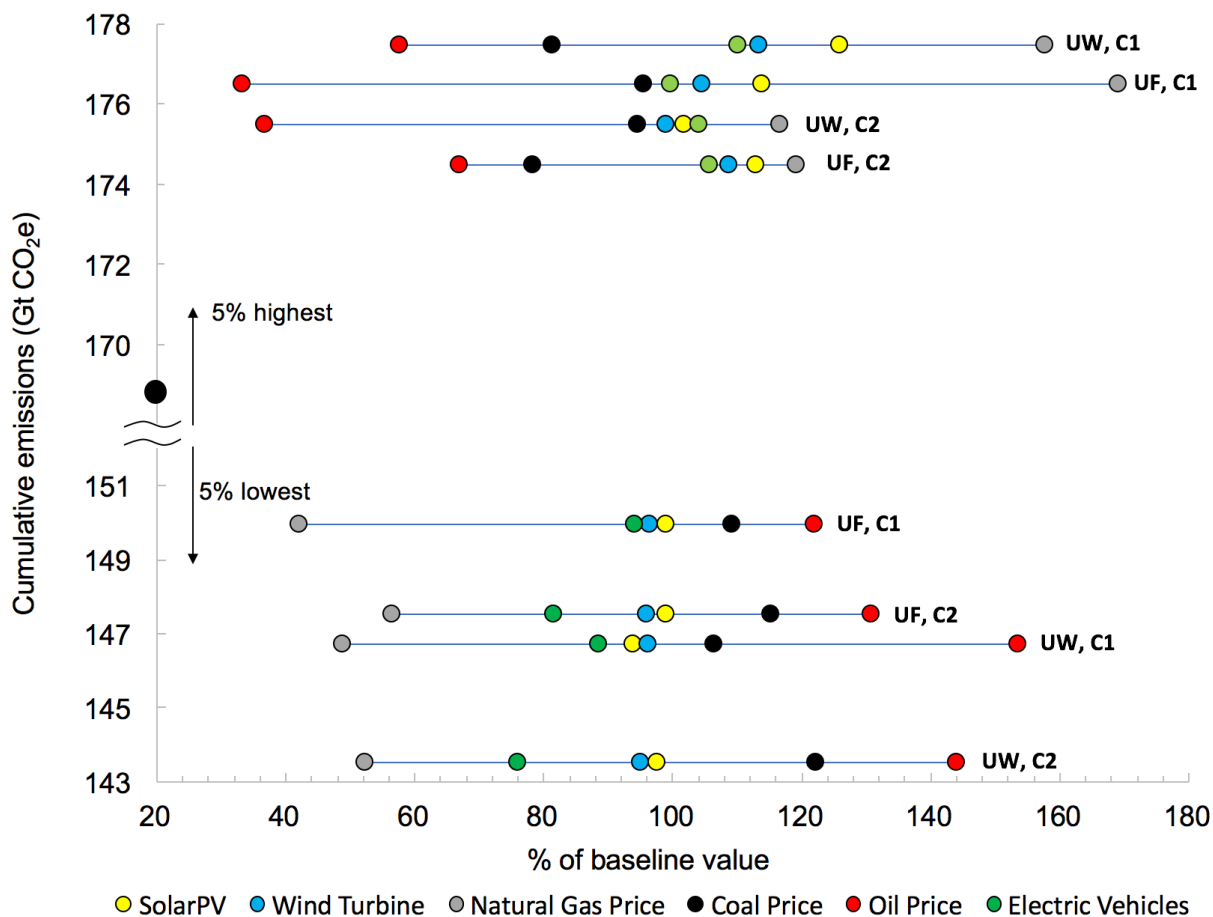

**Figure 6.** Application of *k*-means clustering to the 5% highest and lowest emission runs from the Monte-Carlo simulation for both the Uncertain Fuels ('UF') and Uncertain World ('UW') cases. Each horizontally aligned row represents a single parameter cluster ('C1' or 'C2'), and each colored dot represents the centroid value associated with a specific parameter within the given cluster. The centroid values on the x-axis represent the scaling factors applied to baseline estimates and used in the Monte Carlo simulation; cumulative GHG emissions associated with each cluster are plotted on the y-axis. Cumulative baseline emissions are shown by the black dot on the y-axis.

In the Uncertain Fuels case, both the highest and lowest emissions regimes are characterized by opposing oil and natural gas prices. The centroid values reflect the wider allowable range in natural gas and oil prices (±80%) compared to coal prices and capital costs for alternative technologies (±20%). Because baseline natural gas prices are currently near the lower end of their historical price range, the price reductions required to produce the lowest emissions clusters would be difficult to achieve. Furthermore, since this analysis does not account for correlation between input parameters, we need to consider ex post whether a future with low natural gas prices and high oil prices is plausible. With the advent of shale gas in North American markets, the historically strong correlation between oil and natural gas prices has been weaker since 2007 [53,54]. While there are studies indicating that this decoupling was a temporary phenomenon [55], others show that Henry Hub prices are decoupled from WTI prices [56,57]. Thus, the degree of decoupling between oil and natural gas prices is uncertain, and the assumption here of decoupled prices in the future is at least plausible.

In the Uncertain World case, the centroids associated with the highest emissions clusters include low oil prices and high natural gas prices, with a discernable shift towards lower coal prices and higher capital costs for alternative technologies compared to the base case. We investigated the individual scenarios that comprise the two high emissions clusters, and all are consistent with the centroid values. The centroids associated with the lowest emissions clusters in the Uncertain World case merit careful examination, as they suggest ways in which the lowest emissions pathways can be achieved. In the Uncertain World low emissions clusters, capital cost reductions in electric vehicles coupled with low natural gas prices and high coal prices lead to low electric sector emissions, relatively cheap electricity, and therefore a cost-effective deployment of electric vehicles to supplant gasoline vehicles. The comparison between C1 and C2 in Uncertain World is instructive: relative to C1, the C2 cluster achieves lower emissions with higher coal prices and lower electric vehicle costs. Cluster 2 of Uncertain World achieves the lowest observed emissions with low natural gas prices (52% of baseline), low electric vehicle prices (76% of baseline) coupled with high oil (144% of baseline) and coal prices (122% of baseline). Note that these centroid values do not indicate the relative contribution that each parameter makes to emissions reductions. However, inspection of Figure 6 indicates that the drop in electric vehicle capital cost from Uncertain Fuels Cluster 1 to Uncertain World Cluster 2 is a significant contributor to the 4% drop in cumulative emissions relative the baseline. By contrast, the total drop in cumulative emissions from the baseline to the lowest emissions scenario is approximately 17%.

Thus, electric vehicle deployment is not the dominant factor behind lower emissions, consistent with Babaee et al. [35].

While the *k*-means clustering results strongly suggest the need for low natural gas prices coupled with high oil and coal prices, they obscure some of the underlying variation in the individual scenarios produced by the Monte Carlo simulation. For example, Figure 7 shows the variation in electric sector installed capacity between the baseline and two scenarios drawn from the set of 50 lowest emissions scenarios.

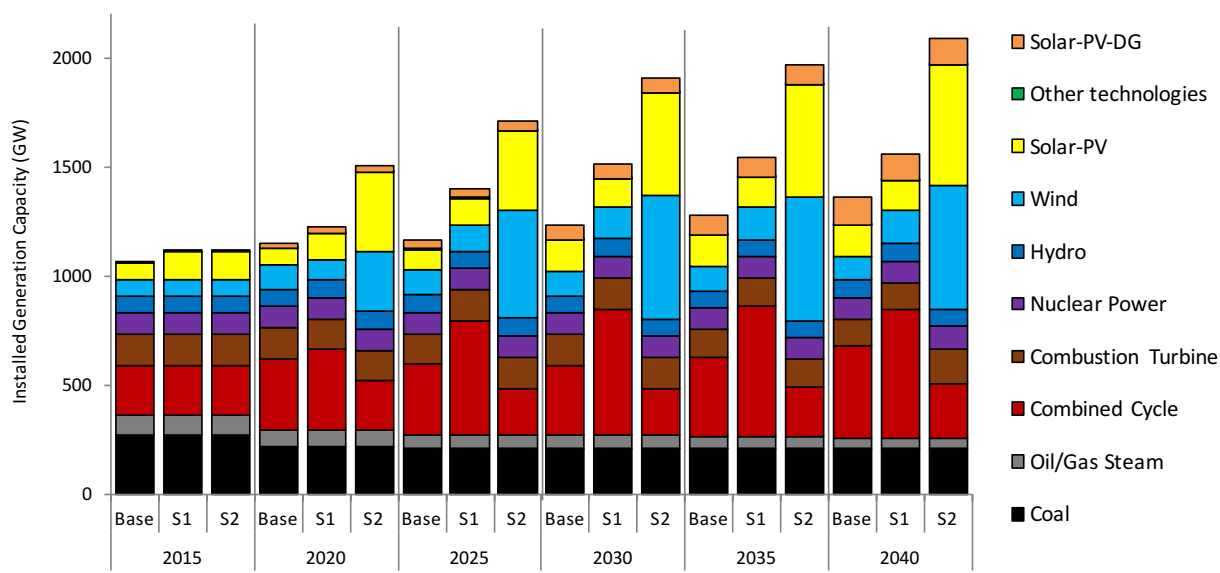

**Figure 7.** Comparison of electric sector capacities in three scenarios: the baseline scenario and two scenarios drawn from the set of 50 lowest emissions scenarios. 'S1' represents a low emissions drawn from Uncertain World Cluster 2 that is consistent with the centroid values shown in Figure 6. 'S2' represents a low emissions scenario drawn from Uncertain World Cluster 1 that shows a result significantly different from the associated centroid values.

The electricity capacity results shown in Figure 7 illustrate the potential diversity in individual scenario results. 'S2' shows a much higher penetration of wind and solar PV compared to either the baseline or 'S1'. The 'S2' scenario achieves among the lowest cumulative greenhouse gas emissions (140 GtCO₂e) with high fossil fuel prices and high combined-cycle turbine cost coupled with capital costs for wind, solar PV, and electric vehicles that are more than 30% below their baseline value.

**Policy Insights and Caveats.** Energy system models are often used to examine a limited number of scenarios that reflect carefully considered states of the world; however, the results often ignore high levels of future uncertainty and can thus be misleading. There is a critical need to introspect energy models to quantify key assumptions, sensitivities, and uncertainties. Real world uncertainty includes a broader array of considerations, such as the prevailing political climate, public acceptance of alternative energy technology, and potential policy actions at the state or regional level that are not captured here. Nonetheless, a careful examination focused on technology cost and performance in a systems context can yield useful insight for policy makers.

Our analysis focuses on techno-economic uncertainty related to fuel prices and technology-specific capital costs, thus providing an indication of how changes in costs can produce different base case outcomes. We do not attempt to model different ranges or correlations among uncertain inputs, which could affect the shape of the emissions distributions shown in Figure 2. Even with a more sophisticated representation of input data, we would not expect a change in the basic insight that technoeconomic uncertainty skews cumulative emissions towards values below the baseline. Our approach here is to conduct the sensitivity analysis with a simplified representation of input data and then examine key relationships ex post for plausibility. This approach leaves open the possibility for new insights. For example, the lowest emissions scenarios rely on low natural gas prices and high oil and coal prices, which led us to consider the degree of price decoupling between these resources. While our assumption of decoupled prices is plausible, future work could test price correlations and their effect on emissions.

Overall, the model results indicate that market forces operating in the absence of new federal climate or energy policy will tend to produce emissions trajectories that remain relatively flat or produce modest reductions: the 2040 emissions range from -23% to +10% of the baseline estimate. By comparison, the 2040 emissions across the AEO 2017 scenarios (without the Clean Power Plan) range from +4% to -5% of the AEO reference scenario [9]. Thus the broader consideration of input uncertainty in this analysis produces a wider range in future emissions, but the range skews towards lower emissions. Our results show consistency with results from Barron et al. [58], where most of the scenarios show relatively flat emissions trajectories in comparison with historical levels. By contrast, Clark et al. [14] and Zhu et al. [59] project higher emissions over the next several decades due to greater reliance on fossil fuels. In our analysis, there are more parameter value combinations that decrease emissions through the deployment of natural gas and renewables than increase emissions through the increased deployment of coal. For perspective, the cumulative difference between the highest and lowest emissions scenario from 2020-2025 is approximately 1.8 times the 2015 emissions level [29], and the same cumulative difference from 2020-2040 grows to nearly 6.6 times the 2015 emissions level [29]. These variations in emissions are significant and illustrate the importance of considering techno-economic uncertainty in future no-policy scenarios. Applying sensitivity techniques that extend beyond conventional scenario analysis can broaden future energy and emissions pathways, and could help inform subsequent policy efforts.

If technology innovation remains low and technology costs track close to their baseline values, then the key tradeoff will be natural gas versus coal utilization in the electric sector. The model results suggest that the continuation of low natural gas prices will lead to additional coal plant retirements, similar to other studies [9,60]. Market forces, policies, and regulations that promote natural gas over coal in the electric sector will lead to lower emissions, though concerted effort is required to minimize upstream methane leakage from natural gas systems [61]. The cluster results (Figure 5) indicate that coal, oil, and natural gas prices as well as capital costs for wind, solar PV, and electric vehicles produce a statistically significant effect on cumulative emissions. The lowest emissions scenarios generally rely on lower natural gas prices and electric vehicle costs in addition to higher oil and coal prices relative to the baseline. The full set of centroids associated with renewable capital costs suggest that they are playing a meaningful role in lowering emissions. For example, Figure 5 indicates that low solar PV costs (12% below the baseline) play a role in achieving cumulative emissions of 160 $GtCO_2e$, which is 5% below the baseline level. Our choice of the 50 scenarios with lowest emissions was illustrative; changing the size of the lowest emissions set could also affect centroid values.

We devised our base case to be conservative. More optimistic assumptions about renewables in the baseline could shift the cost threshold at which renewables are deployed at large scale. In addition, our model does not include the EPA Clean Power Plan[62]. While the collective requirement under state-level renewable portfolio standards is included, we did not explicitly model emissions caps in California or the Northeastern states under RGGI. These existing policies, combined with additional state-level efforts to reduce emissions and increase the deployment of renewables, could produce significant GHG reductions beyond those estimated here. Our analysis indicates that energy market forces, operating in the absence of significant new policy, will hold emissions close to current levels or produce modest reductions. While it is heartening that a hiatus in federal energy and climate policy will not produce a dramatic rise in emissions, aggressive policy action will be required to produce the level of GHG reductions required to avoid the worst effects of climate change.

## ASSOCIATED CONTENT

**Supporting Information:** Description of model input data, baseline scenario results, and mathematical formulation of Method of Morris and k-means clustering algorithm.

## ACKNOWLEDGMENTS

This material is based upon work supported by the National Science Foundation under Grant No. CCF-1442909 and by a graduate fellowship awarded to the first author by the Department of the Interior Southeast Climate Adaptation Science Center. Any opinions, findings, and conclusions, or recommendations expressed in this material are those of the authors and do not necessarily reflect the views of the National Science Foundation.

**TOC Art**

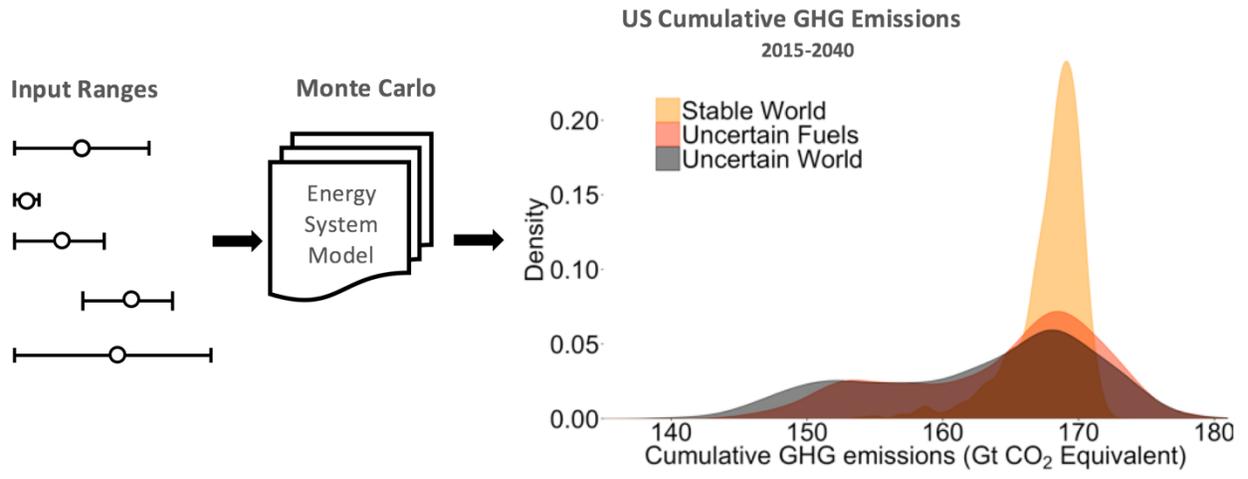

**US Energy-Related Greenhouse Gas Emissions in the Absence of Federal Climate Policy**

Hadi Eshraghi[1], Anderson Rodrigo de Queiroz[2], Joseph DeCarolis[1,*]

[1] Department of Civil, Construction, & Environmental Engineering, North Carolina State University, Raleigh, NC 27695-7908.
[2] School of Business, NC Central University, Durham, NC 27707.
[*]Corresponding author email: jdecarolis@ncsu.edu

# Supporting Information

## Table of Contents

**Summary of Supporting Information:**
21 pages, including coversheet

**Tables**
**S1.** Capacity factor of wind and solar technologies
**S2.** Minimum renewable electricity production (%) based on existing state-level RPSs
**S3.** Investment costs for solar PV (2015 $/kW)
**S4.** Groups of the input parameters for the sensitivity analysis

**Figures**
**S1.** Reference energy system showing high-level energy commodity flows between sectors
**S2-S9.** Sector-specific baseline results compared with the Annual Energy Outlook 2017

## 1. Temoa input database description

The model time horizon spans 2015 to 2040, with 5-year time periods. To represent seasonal and diurnal variations in energy supply and demand, the model must perform energy commodity balances across a set of time slices that represent different combinations of seasons and times of day. In the input database used in this analysis, we represent three seasons (summer, winter, intermediate) and four times of times of day: AM (6AM - 12PM), peak (12PM - 3PM), PM (3PM - 9PM), and night (9PM - 6AM) based on the US EPA national database [1]. The US energy system is modeled as a single region, and a 5% social discount rate is used to bring future costs back to the present. All costs are adjusted to 2005 US dollars.

Fuel prices are exogenous to the model and are extracted from the 2017 Annual Energy Outlook (AEO) base case results that assume the EPA Clean Power Plan is not implemented. While we make the simplifying assumption that fuel prices are not responsive to endogenous changes within the model, retrospective analysis indicates that there is considerable uncertainty associated with the projection of future fuel prices [2].

The end-use sectors include demand technologies that convert secondary energy carriers (e.g., electricity, natural gas, liquid fuels) into useful energy services (e.g., space heating, space cooling, vehicle miles traveled). These energy service demands are specified exogenously and are drawn from the most recent version of the EPA MARKAL database [1]. For example, the residential sector includes demands for space heating, space cooling, water heating, freezing, refrigeration, lighting, and miscellaneous electricity for appliances. A schematic of the Temoa input dataset is shown in Figure S1.

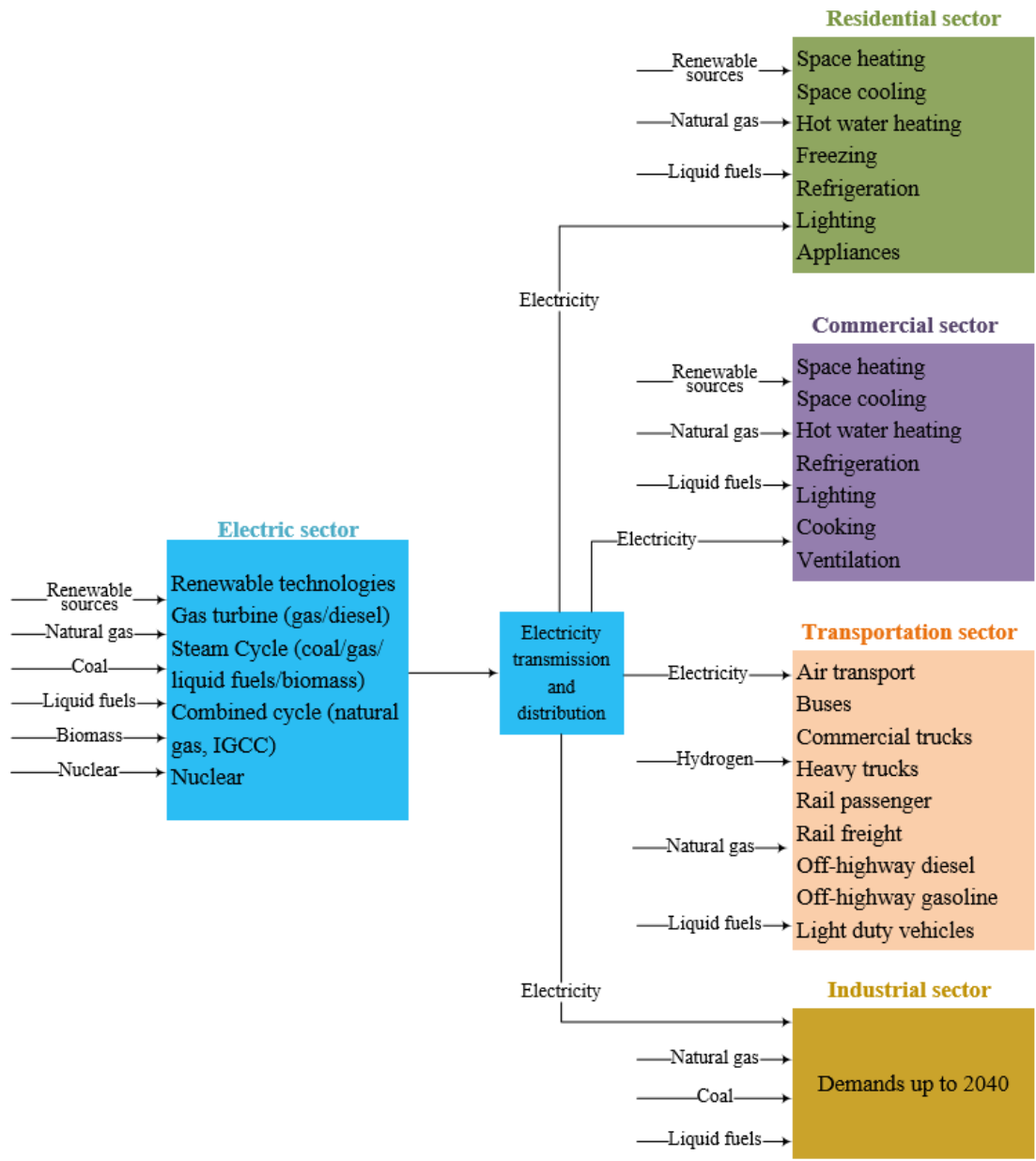

**Figure S1.** Reference energy system showing high-level energy commodity flows between sectors within the model. Energy commodities (other than electricity) are specified exogenously and drawn from EPA MARKAL database [1]. Labels within the rightmost sector boxes represent the specific end-use demands that must be met.

Below we expand on our electric sector representation, use of hurdle rates, and baseline policy specification within in the model.

## 1.1. The electric sector

The electricity sector modeled in this analysis includes a representation of existing and new generation technologies. Thermal power plants include coal-fired steam, integrated gasification combined-cycle (IGCC) with and without carbon capture and storage (CCS), oil-based steam plants, natural gas steam plants, open cycle and combined-cycle natural gas turbines with and without CCS, and light water nuclear reactors. Renewable sources include conventional hydro, solar photovoltaics, concentrating solar thermal, wind, biomass steam, waste-to-energy, and geothermal. In addition to these electric generating technologies, the model represents air pollution retrofit technologies for $NO_x$ removal, including low $NO_x$ burners (LNB), selective catalytic reduction (SCR), and selective non-catalytic reduction (SNCR). In addition, flue gas desulfurization (FGD) can remove $SO_2$ associated with coal-fired generation. Techno-economic data used to parameterize technologies are gathered from the most recent version of the EPA MARKAL model [1] as well as the Annual Energy Outlook (AEO) input assumptions [3].

Wind and solar capacity factors by model time slice are given in Table S1. The specified capacity factors of wind and solar are drawn from the average values in the EPA MARKAL model [1] and NREL's Solar Power Data for Integration Studies [4] respectively. In order to appropriately reflect the regionally varying nature of wind resources throughout the country, we defined three classes of wind technologies that vary by their capacity factor and upper bound availability. Wind resources include Class 4, Class 5 and Class 6 with upper bounds of 240, 387 and 106 gigawatts, respectively [1].

**Table S1.** Capacity factor of wind and solar technologies

| Time Slice | Fraction of Year | Wind Class 6 | Wind Class 5 | Wind Class 4 | Solar Thermal | Solar PV (utility scale) | Solar PV (rooftop) |
|---|---|---|---|---|---|---|---|
| Winter-AM* | 0.082 | 0.515 | 0.464 | 0.412 | 0.46 | 0.34 | 0.24 |
| Winter-Peak* | 0.003 | 0.515 | 0.464 | 0.412 | 0.46 | 0.43 | 0.31 |
| Winter-PM* | 0.096 | 0.515 | 0.464 | 0.412 | 0.46 | 0.34 | 0.24 |
| Winter-Night* | 0.153 | 0.372 | 0.335 | 0.298 | 0.23 | 0 | 0 |
| Summer-AM* | 0.098 | 0.398 | 0.358 | 0.318 | 0.46 | 0.42 | 0.30 |
| Summer-Peak* | 0.003 | 0.398 | 0.358 | 0.318 | 0.46 | 0.56 | 0.4 |
| Summer-PM* | 0.109 | 0.398 | 0.358 | 0.318 | 0.46 | 0.53 | 0.38 |
| Summer-Night* | 0.125 | 0.286 | 0.258 | 0.229 | 0.23 | 0 | 0 |
| Intermediate-AM* | 0.082 | 0.515 | 0.463 | 0.412 | 0.46 | 0.39 | 0.30 |
| Intermediate-Peak* | 0.003 | 0.515 | 0.463 | 0.412 | 0.46 | 0.53 | 0.38 |
| Intermediate-PM* | 0.109 | 0.515 | 0.463 | 0.412 | 0.46 | 0.53 | 0.38 |
| Intermediate-Night* | 0.138 | 0.398 | 0.335 | 0.318 | 0.23 | 0 | 0 |
| Country Average** | - | 0.42 | 0.37 | 0.34 | | 0.25 | 0.16 |
| Upper bound on installed capacity (GW) | | 106 | 387 | 240 | | | |

* AM: 6AM to 12PM, Peak: 12PM to 3PM, PM: 3PM to 9PM, Night: 9PM to 6AM
** Weighted average by time-slice.

## 1.2. Hurdle rates

Hurdle rates represent the technology-specific discount rates used to amortize capital costs, and can be used to represent non-economic costs such as time preference, risk, and uncertainty [5]. Without hurdle rates on new technologies in the residential and commercial sectors, the majority of existing end-use technologies are retired in the first time period and replaced with new capacity. For simplicity, Temoa does not consider the remaining capital payments on existing technology, which often makes it cost-effective to simply replace older, less efficient vintages with new ones. To remedy this issue, we assigned a uniform hurdle rate of 30% to all the new technologies in the residential and commercial sector. This rate is high enough to keep existing technologies active until they reach the end of their useful lifetimes. Since the hurdle rate is uniform for all new technologies, it doesn't incentivize one technology over another. Adhering to the same logic in the electric sector, all electric generating technologies use a uniform 6% rate, which is the rate for renewable and natural gas-fired technologies in AEO 2017 [3].

We used a uniform 10% hurdle rate for all the alternative vehicles in the light duty vehicle sector and 5% for conventional internal combustion engines. All the other technologies in the transportation sector have 5% hurdle rates. After we ran the model with this setting, we observed zero deployment of alternative light duty vehicles in the base case. For simplicity, we assume that alternative vehicles (excluding ethanol-flex fuel vehicles) must obtain at least a 10% market share by 2025 and thereafter. This approximates the adoption of alternative vehicles by early adopters who are less risk averse, and also produces results that roughly match the AEO 2017 base case. The assumed 10% hurdle rate is relatively low; surveys have estimated hurdle rates associated with alternative vehicle purchases in the range of 20-50% [6,7,8].

### 1.3. Policies

State-level renewable portfolio standards (RPSs) are included in our single region dataset in the form of minimum production shares from renewables. Total electricity generation from renewables are drawn from the AEO 2018 input assumptions [3] and are then divided by the total electricity generation from AEO 2018 [2] to obtain the minimum shares of renewable electricity. Data regarding RPSs are shown in Table S2. Since the goal of an RPS is to increase the deployment of new renewable technologies, most states restrict existing hydro plants from counting towards the RPS target. Drawing the eligible hydro plants under an RPS from EPA MARKAL [1], it is assumed that only 28% of the hydro capacity qualifies for the minimum production level constraint [1].

**Table S2.** Minimum renewable electricity production (%) based on existing state-level RPSs

|  | 2020 | 2025 | 2030 | 2035 | 2040 |
|---|---|---|---|---|---|
| Renewable share | 16 | 18 | 20 | 20 | 20 |

Federal incentive programs, including the Investment Tax Credit (ITC) [9] and Production Tax Credit (PTC) [10], are also taken into account based on their current values and scheduled sunsets. The ITC applies to solar technologies and the PTC to new wind turbines. For utility scale solar PV, a 30% reduction in investment costs for the first period (2015) is assumed, and the credit is reduced to 10% beginning in the 2020 time period. Residential rooftop solar PV is subject to the same federal program, except no credit will be awarded after 2021 [9]. Furthermore, solar PV is assumed to incur annual cost reductions as a result of technological learning, based on approximation of AEO learning rates [1]. AEO 2018 endogenously specifies a 13% cost decline for each doubling of capacity [3]. Looking at their utility-scale solar PV projection capacities in 2040, we assume that PV capacity will double twice between 2018-2040. These two doublings correspond to 25% decline in 2040, which adds up to 35% due to ITC. The cost of utility-scale solar PV is 1850 as per AEO 2018 $/kW$_{ac}$ [3] and with a 35% decline reaches to 1202.5 $/kW$_{ac}$. Rooftop solar PV costs are directly drawn from the residential demand module of AEO 2018 input assumptions [3]. The resultant baseline investment costs for solar PV technologies are listed in Table S3.

**Table S3.** Investment costs for solar PV (2015 $/kW)

| Technology | 2015 | 2020 | 2025 | 2030 | 2035 | 2040 |
|---|---|---|---|---|---|---|
| Utility scale solar PV | 1855 | 1665 | 1549 | 1433.75 | 1318.1 | 1202.5 |
| Rooftop solar PV | 2582 | 3221 | 2623 | 2027 | 1858 | 1689 |

For wind, a production tax credit of 0.014 $/kWh is specified for the plant with construction beginning in 2015-2019. This value is the average of the credits mentioned in the Consolidated Appropriation Act of 2016 passed in December 2015 [10].

## 2. Results from the baseline scenario

### 2.1. End-use sectors

Figures S2 and S3 show energy consumption in the residential and commercial sectors, respectively. For comparison, the same results from the Annual Energy Outlook (AEO) 2017 Reference scenario without CPP [11] are also shown.

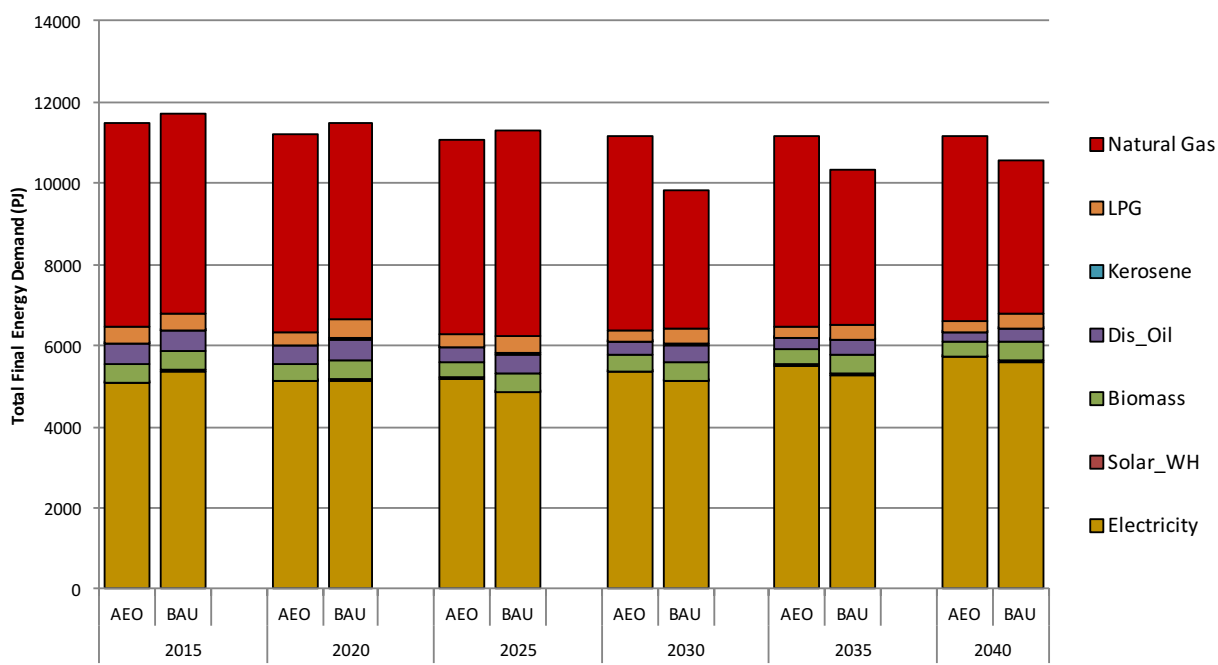

**Figure S2.** Residential sector baseline energy consumption in AEO (left) and this analysis (right).

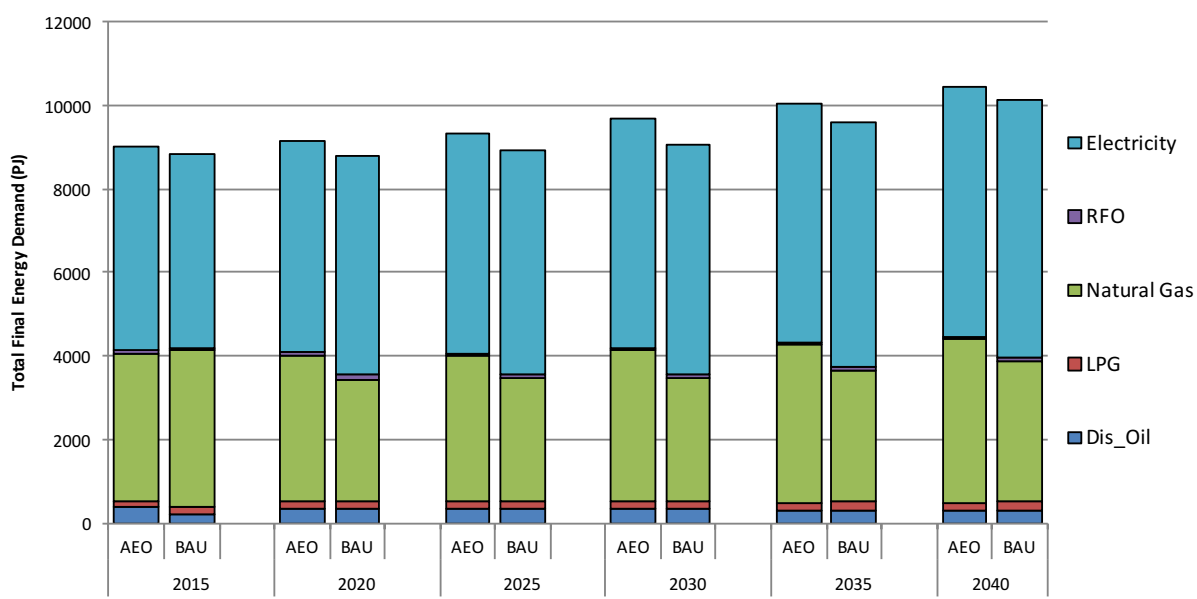

**Figure S3.** Commercial sector baseline energy consumption in AEO (left) and this analysis (right).

In both sectors, a shift to more efficient technologies is observed in the Temoa baseline scenario compared with AEO. Less delivered energy is consumed in our model compared to AEO in order to meet approximately the same end-use services.

Figure S4 shows the fuel consumption in the transportation sector. When compared to the AEO results, the Temoa baseline scenario shows less consumption of E10 (a gasoline blend containing 10% ethanol) and diesel from 2025 to 2040. While our vehicle cost and performance estimates are aligned with AEO, our model is less complex and tends toward more efficient vehicles in order to minimize cost.

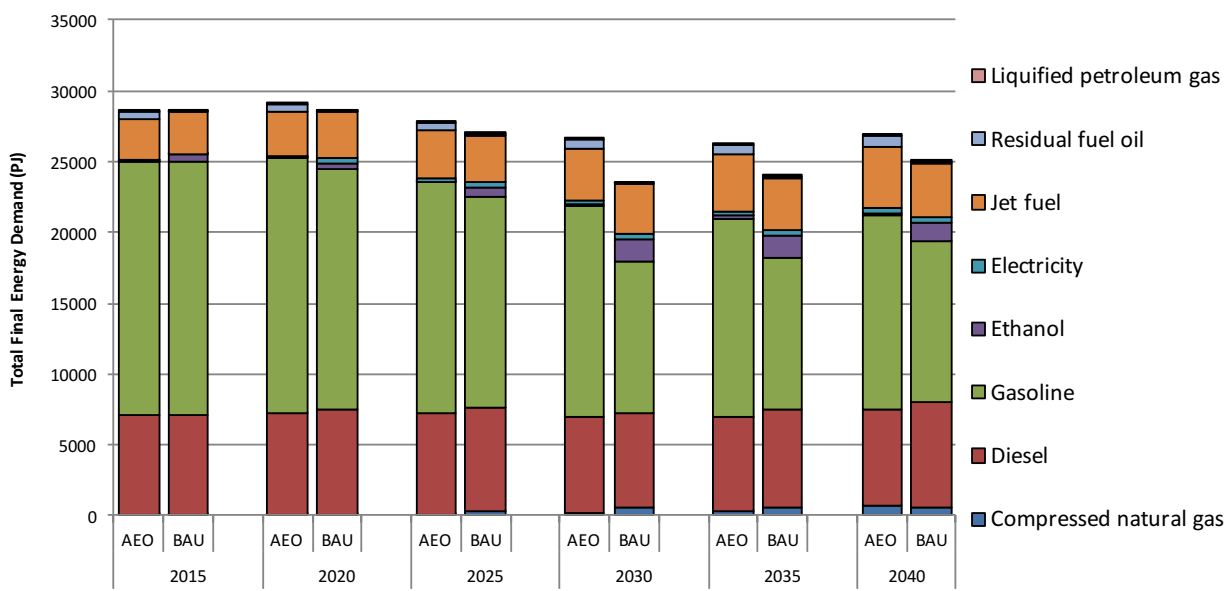

**Figure S4.** Transportation sector baseline energy consumption in AEO (left) and this analysis (right).

The technological mix in the light duty vehicle sector is shown in Figure S5. The deployment of electric vehicles is largely due to the constraint that alternative vehicle technologies must make up a minimum 10% market share. Among alternative vehicle technologies, battery electric vehicles are the most cost-effective. Regarding E85, the prevailing AEO prices make ethanol vehicles cost-competitive, however, an upper bound drawn from Renewable Fuel Standard on ethanol availability [12] limits E85 consumption in the transportation sector.

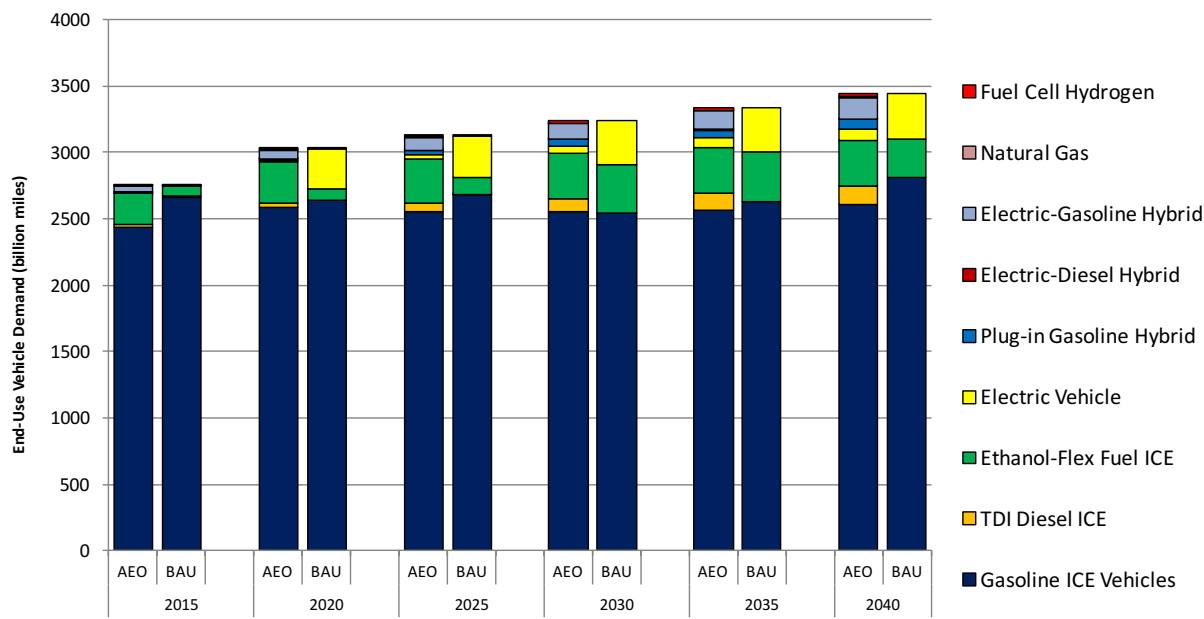

**Figure S5.** The baseline light duty vehicle technology mix in AEO (left) and this analysis (right).

The industrial sector is another major end-use sector represented in Temoa. The current modeling framework of this sector however does not allow flexibility regarding technological and fuel mix change, and simply follows the AEO base scenario projections. As is noted by Barron [14], we recognize that this modeling framework hinders our ability to evaluate technological shifts and policies implementation in the industrial sector.

## 2.2. Electric sector

Base case electric sector results are presented in Figure S6.

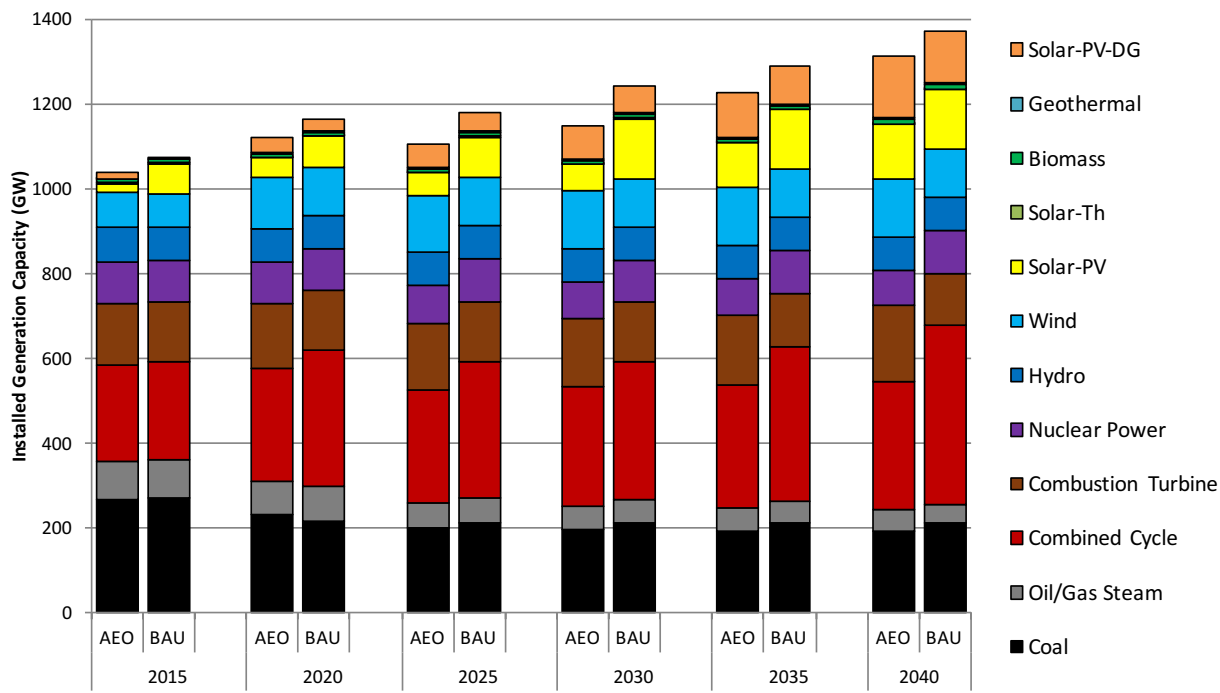

**Figure S6.** Baseline electric sector installed capacity in AEO (left) and this analysis (right).

The Temoa baseline scenario builds more advanced combined cycle plants and utility-scale solar PV and slightly less wind than AEO. The National Energy Modeling System (NEMS), which is used to produce AEO, uses a zip code-level econometric model to project rooftop solar PV adoption [15]. Since our model does not capture all the dynamics relevant to residential solar PV deployment, we adopt the AEO 2018 base case capacities of rooftop solar PV in the form of minimum capacity constraints. Temoa deploys utility-scale solar PV at a higher rate compared to AEO. However, this deployment is largely due to the aggregation of state-level RPSs. Figure S7 shows electric sector capacities in the absence of RPS.

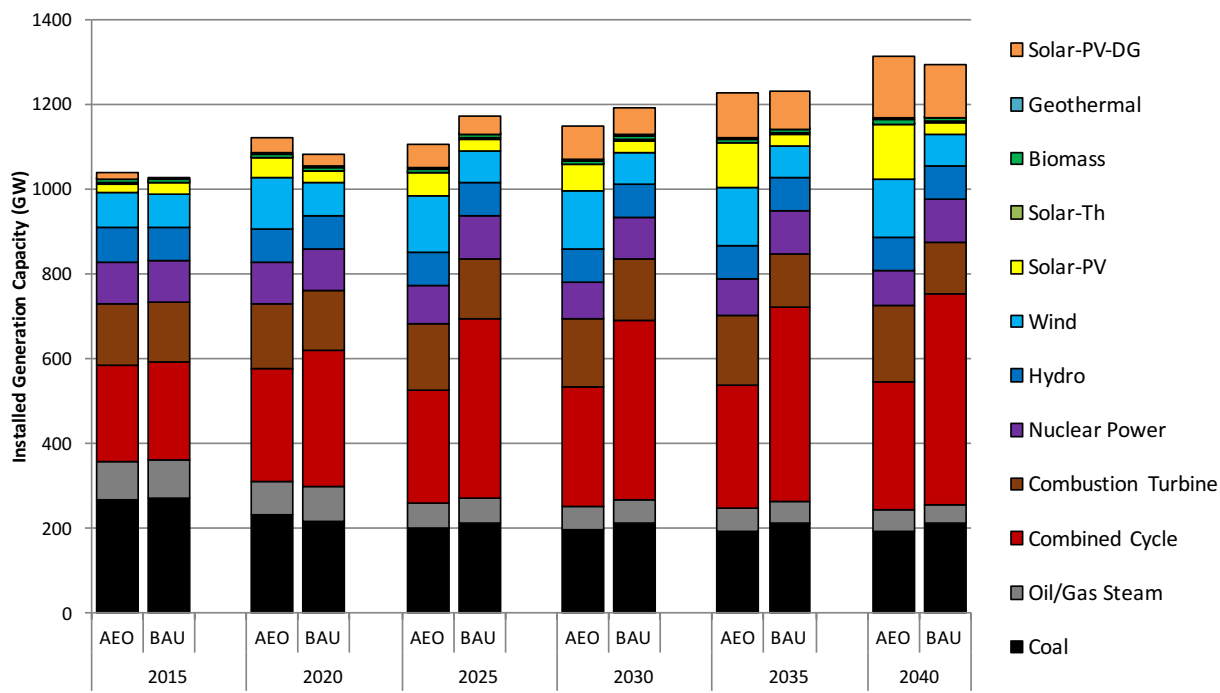

**Figure S7.** Baseline electric sector installed capacity in AEO (left) and this analysis without the RPS (right).

As is shown in Figure S8, both the Temoa and AEO baseline scenarios project nearly similar levels of electricity demand. Coal retains a significant share of electricity generation through the model time horizon in both models. In the first model time period, 2015-2019, existing capacity in 2015 is fixed, but new wind and solar PV capacity in the electric sector are allowed, given their rapid annual growth in recent years.

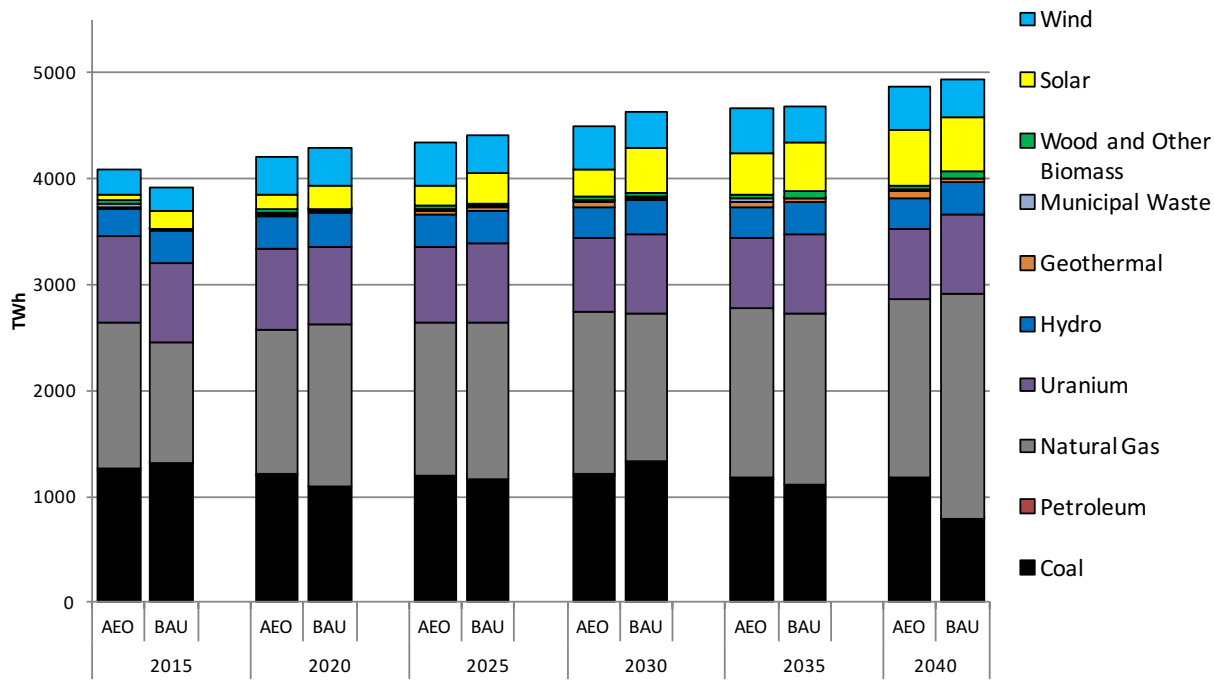

**Figure S8.** Baseline electric sector generation in AEO (left) and this analysis (right).

## 2.3. Emissions

Temoa estimates emissions of $CO_2$, $SO_2$ and $NO_x$ from direct fuel combustion in the energy sector using EPA emission factors [1]. In addition, $CO_2$ equivalent emissions from the upstream natural gas production system (both combustion-related $CO_2$ emissions and fugitive methane emissions) are estimated using actual emissions in 2015. According to EPA [16], natural gas production in 2015 resulted in 42 million metric tons of combustion-related $CO_2$ and 162.4 million tonnes of $CO_2e$ methane, assuming a global warming potential of 25 for methane. Total US natural gas consumption in 2015 was approximately 28,000 PJ, yielding an emissions factor of 7.3 ktons $CO_2$-e /PJ for natural gas extraction, processing, storage, transportation and distribution.

Figure S9 shows energy-related $CO_2$ emissions from all sectors. The $CO_2$ emissions nearly level off during 2030-2040 and experience a modest 5% drop during 2015-2030. While electricity-related $CO_2$ emissions remain at their current levels (approximately 1800 million metric tons), the transportation sector experiences the most $CO_2$ reduction due to the improvement in vehicle fuel economy and the partial switch to electric vehicles in the light duty sector.

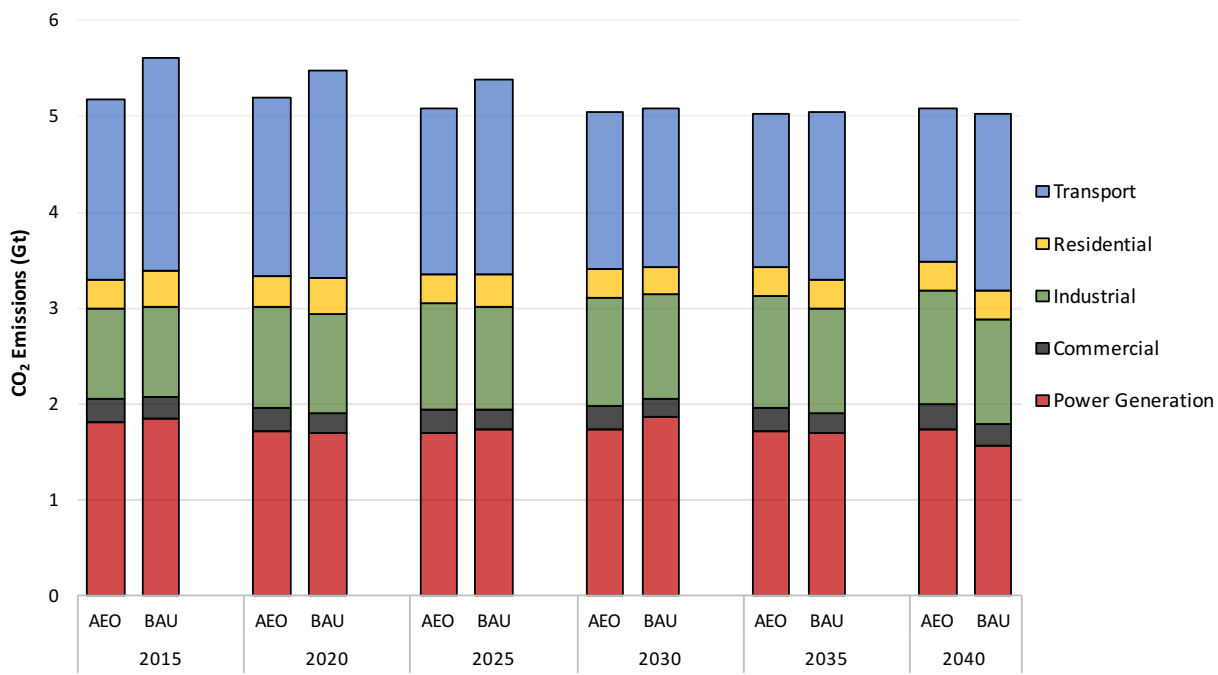

**Figure S9.** Sectoral $CO_2$ emissions through 2040 in the base case. While $CO_2$ equivalent emissions associated with methane leakage are included in the Temoa database, only direct $CO_2$ combustion-related emissions are shown here for comparison with AEO.

## 3. Method of Morris

Method of Morris works by calculating a number of incremental ratios called Elementary Effects (EE), sampled on a grid representing the input parameter space [16,17]. Assuming $k$ input parameters, the sensitivity experiment is based on one-at-a-time changes to each parameter $x_i, i = 1, \dots, k$, which is assumed to vary across $p$ levels in the discretized input space $\Omega$. As a standard practice, this input space is assumed to be the unit hypercube, in which $k$ independent inputs are uniformly distributed across $p$ discrete levels. The elementary effect of the $i$th input ($EE_i$) is defined by equation (1):

$$EE_i = \frac{f(x_1, x_2, \ldots, x_i + \Delta, \ldots, x_k) - f(x_1, x_2, \ldots, x_i, \ldots, x_k)}{\Delta} \qquad (1)$$

where $\Delta$ represents a perturbation to $x_i$ and can assume arbitrary values $\{\frac{1}{p-1}, \frac{2}{p-1}, \ldots, 1 - \frac{1}{p-1}\}$ and $x = (x_1, x_2, \ldots, x_i, \ldots, x_k)$ is any selected point in the $k$-dimensional $p$-level grid ($\Omega$). Note that $x + u_i \Delta$ is still in $\Omega$, where $u_i$ is a $k$-dimensional vector with zero elements for all its components but unity for the $i$th one.

To implement the method, the elementary effects associated with each input factor must be sampled. Morris[16] proposes a sampling strategy based on developing $N$ different trajectories in the $k$-dimensional input parameter space, $\Omega$, each of which consists of $(k + 1)$ points. Each trajectory generates a total number of $k$ elementary effects, one per each input. Thus, a total number of $N \times (k + 1)$ model executions are needed.

Even though the number of trajectories $N$ is exogenously defined, it interacts with the value chosen for $p$. A higher $p$ means a higher resolution $\Omega$, and consequently, to better cover $\Omega$, a larger $N$ is required. Saltelli et al.,[18] suggests using more than 10 trajectories when $p = 4$. The mean and standard deviation associated with the resultant distribution of elementary effects, $F_i$, reveals information about the $i$th input. A large value for the mean indicates that the input has a high influence on function $f$, and a large measure for the standard deviation suggests that the input interacts with other inputs, has a nonlinear effect, or both.

Campolongo et al.[17], as an extension to the original Morris methodology[16], proposed that the average of the absolute values of elementary effects, $\mu_i^*$, is in fact a better measure of sensitivity because it rules out the possibility of elementary effects with opposing sign cancelling one another out. In this modification, $\mu_i^*$ as defined by equation (2), is a measure of the expected variance of function $f$ when only the $i$th input can change, given the interactions with other inputs:

$$\mu_i^* = \frac{\sum_{j=1}^{N} |EE_i^j|}{N} \qquad\qquad (2)$$

where $EE_i^j$ is the elementary effect associated with the $i$th input along the $j$th trajectory and $N$ is the number of the trajectories. We use the approach suggested by Campolongo et al.[17], and Figure 1 of the main manuscript presents $\mu_i^*$ divided by 2015 emissions.

In this study, $\Omega$ consists of 41 parameter groups (Table S4), and the range associated with each parameter is ±20% of its baseline value. A single trajectory therefore consists of 42 points, and N=25 trajectories are created. Because Temoa is dynamic, parameter values can vary by model time period. As a result, a single parameter indexed by time period constitutes a single parameter group. For example, natural gas prices over the model time horizon constitute a single group, and thus the natural gas price trajectory is uniformly shifted up or down within Method of Morris, rather than allowing prices to shift randomly from one time period to the next. Grouping ensures consistent trajectories for capital costs and fuel prices, and reduces the computational effort. To conduct this analysis, we make use of SALib[19], an open source Python library, which includes a complete implementation of the Method of Morris.

**Table S4.** Groups of the input parameters for the sensitivity analysis

| Investment cost parameter (35 groups) | | Fuel price parameter (6 groups) |
|---|---|---|
| Group | No. of Included Technologies | |
| Boiler for space heating/water heating | 5 | Biomass price |
| Furnace for space heating | 10 | Coal price |
| Radiative space heating | 10 | Natural gas price |
| Heat pumps | 11 | Petroleum product prices |
| Fluorescent lighting | 10 | Hydrogen price |
| Geothermal heat pumps | 2 | Uranium price |
| HID lighting | 2 | |
| HID-LED lighting | 3 | |
| LED lighting | 4 | |
| Resistive lighting | 2 | |
| Induction cooking | 2 | |
| Convection cooking | 2 | |
| Freezing and refrigeration | 4 | |
| Solar water heating | 1 | |
| Ventilation | 1 | |
| Coal steam plant | 1 | |
| Combined cycle plants | 1 | |
| Combustion turbine plants | 1 | |
| Carbon Capture and Storage (CCS) | 1 | |
| Integrated Gasification Combined Cycle (IGCC) | 1 | |
| Nuclear plants | 1 | |
| Solar PV | 2 | |
| Solar thermal | 1 | |
| Wind turbine | 3 | |
| Geothermal | 1 | |
| Compressed Natural Gas (CNG) vehicles | 6 | |
| Diesel engines | 18 | |
| Diesel engine railroads | 6 | |
| Electric railroad | 3 | |
| Conventional internal combustion engines | 21 | |
| Hybrid vehicles (gasoline-based) | 14 | |
| Hybrid vehicles (diesel-based) | 8 | |
| Plugin hybrid vehicles (gasoline as the main fuel) | 22 | |
| Electric vehicles | 3 | |
| Fuel cell vehicles | 5 | |

Changing the input parameters uncertainty ranges used by Method of Morris can affect the parameter sensitivity rankings, which in turn affect the Monte Carlo simulation. As a side case, we developed the upper bound of input ranges based on Muratori et al[20]. This led to the removal of wind capital cost and the addition of coal steam investment cost from the Method of Morris rankings. However, after performing the Monte Carlo simulation, we found that this change did not produce a significant change in the shape of the emissions distributions in Figure 2 of main manuscript. In another test similar to the -+20% input parameter range, we let the input parameters vary by -+40%. This produced the same top ten parameters but this time oil prices ranked second while the relative rank of the rest of the parameters stayed the same. The results support our general insight that the system uncertainty tends to skew emissions towards lower values.

## 4. K-means clustering

K-means clustering partitions a group of $n$ observations $O_1, O_2, \ldots O_n$ into a set of $k \ (< n)$ clusters: $S = \{S_1, S_1, \ldots, S_k\}$. Given $n$ observations, each observation $O_i$, $i = 1, \ldots, n$ can have $m$ attributes that define the position of that observation in $m$-dimensional space. The goal of $k$-means clustering is to find the centroids associated with each of the $k$ clusters $C_i$, $i = 1, \ldots, k$ in the $m$-dimensional space, such that the Euclidean distance $D$ in Equation (3) is minimized:

$$D = \sqrt{\sum_{i=1}^{k} \sum_{O_j \in S_i} (O_j - C_i)^2} \tag{3}$$

Where $S_i$ belongs to the set of clusters $S = \{S_1, S_1, \ldots, S_k\}$ and $O_j, C_i \in \mathbb{R}^m$.

Minimizing equation (3) is computationally difficult because it is NP-hard[21]. Instead of solving equation (3) directly, heuristic algorithms are used to find the optimal centroids. The most common approach, and the one used here, is Lloyd's algorithm [22]. The algorithm works iteratively by: (1)

initializing the centroids $C_i$, $i = 1, \dots, k$; (2) assigning the $n$ observations to the closest cluster; (3) updating the centroids of each of the clusters with the observations assigned to that cluster; and (4) repeating Steps 1 to 3 until there is no difference in the value of the centroids.